\documentclass[12 pt]{article}
\usepackage{graphics}
\usepackage{amssymb}
\usepackage[driver]{graphicx}

\newcommand{\QED}{\mbox{\rule[-1.5pt]{6pt}{10pt}}}

\newcommand{\N}{\mathbb{N}}
\newcommand{\R}{\mathbb{R}}

\newcommand{\OO}{{\cal O}}

\newtheorem{claim}{Claim}[section]
\newtheorem{theorem}[claim]{Theorem}
\newtheorem{proposition}[claim]{Proposition}
\newtheorem{lemma}[claim]{Lemma}

\begin{document}

\title{Bound states in point-interaction star graphs}
\author{P.~Exner$^{a,b}$ and K.~N\v{e}mcov\'{a}$^{a,c}$}
\date{}
\maketitle
\begin{quote}
{\small \em a) Department of Theoretical Physics, Nuclear Physics
Institute, \\ \phantom{e)x}Academy of Sciences, 25068 \v Re\v z,
Czechia \\
 b) Doppler Institute, Czech Technical University, B\v{r}ehov{\'a} 7,\\
\phantom{e)x}11519 Prague, Czechia \\
 c) Faculty of Mathematics and Physics, Charles University,\\
\phantom{e)x}V~Hole\v{s}ovi\v{c}k\'ach~2, 18000 Prague, Czechia
\\
 \rm \phantom{e)x}exner@ujf.cas.cz, nemcova@ujf.cas.cz}
\vspace{8mm}

\noindent {\small We discuss the discrete spectrum of the
Hamiltonian describing a two-dimensional quantum particle
interacting with an infinite family of point interactions. We
suppose that the latter are arranged into a star-shaped graph with
$N$ arms and a fixed spacing between the interaction sites. We
prove that the essential spectrum of this system is the same as
that of the infinite straight ``polymer'', but in addition there
are isolated eigenvalues unless $N=2$ and the graph is a straight
line. We also show that the system has many strongly bound states
if at least one of the angles between the star arms is small
enough. Examples of eigenfunctions and eigenvalues are computed
numerically.}
\end{quote}


\section{Introduction}

Graph-type systems are used in quantum mechanics for a long time
\cite{RS}, but only in the last decade they became a subject of an
intense interest -- cf.~\cite{KS} and references therein. Among
various graph geometries, star graphs were investigated from
different point of view. Recall, for instance, a natural
generalization of the weak-coupling analysis for one-dimensional
Schr\"odinger operators \cite{E1}, signatures of quantum chaos
found recently in stars with finite nonequal arms \cite{BK}, etc.

From the mathematical point of view Schr\"odinger operators on
graphs are easy to deal with, because they represent systems of
Sturm-Liouville ODE's coupled through boundary conditions at the
graph vertices. This is due to the assumption that the
configuration space of the system is just the graph. From the
physical point of view, this is certainly an idealization. One of
the most common applications of graph models is a description of
various mesoscopic systems like quantum wires, arrays of quantum
dots, etc. In reality their boundaries are finite potential steps,
and therefore the particle can move away from the prescribed area,
even if not too far because the exterior of such a graph is a
classically forbidden region.

There are various ways how to model such ``leaky'' graphs. One can
use Schr\"odinger operator with a Dirac measure potential
supported by the graph -- see \cite{BT, EI} and references
therein. Here we consider another, in a sense more singular model
where the graph is represented by a family of two-dimensional
point interactions. Its advantage is that such a model is solvable
because (the discrete part of) the spectral analysis is reduced
essentially to an algebraic problem. Two-dimensional
point-interaction Hamiltonians were studied by various authors --
references can be found in the monograph \cite{AGHH}.
Nevertheless, relations between spectral properties of such
operators and the geometry of the set of point-interaction sites
did not attract much attention. Here we are going to fill this gap
partly by discussing an example of a point-interaction star graph.

The model is described in the following section. Next, in
Section~3, we show that the essential spectrum is given by the
structure of each graph arm at large distances and thus it
coincides with that of an infinite straight ``polymer''
\cite[Sec.~III.4]{AGHH}. More surprising is the fact that a star
graph has a nonempty discrete spectrum, with the exception of the
trivial case when the graph is a straight line. This is proved in
Section~4 where we also show that there are geometries which give
rise to numerous strongly bound states. In the final section we
present numerically computed examples showing eigenvalues and
eigenfunctions for various graph configurations. Of course, the
discrete spectrum is not the only interesting aspect of these
Hamiltonians. One can ask about the scattering, perturbations
coming either from changes in the geometry or from external
fields, etc. We leave this questions to a future publication.


\section{Formulation of the problem}

For a given integer $N \geq 2$, consider an $(N\!-\!1)$-tuple of
positive numbers $\beta:=(\beta_1,\ldots,\beta_{N-1})$ such that
$\sum_{j=1}^{N-1} \beta_j < 2\pi $ and denote
$\vartheta_j:=\sum_{i=1}^j \beta_i$ and $\vartheta_0:=0$. Then one
can define the set
 $$
Y=\bigcup_{j=0}^{N-1}\{(n l \cos(\vartheta_j),n l
\sin(\vartheta_j))\} _{n\in\N} \cup (0,0),
 $$
where $l>0$ is a given distance which has the meaning of the
spacing of points at each ``arm'' of $Y$.

The object of our study is a two-dimensional Hamiltonian, which we
denote as $H_N(\alpha,\beta)$, with a family of point interactions
supported by the set $Y$ having the same ``coupling constant''
$\alpha$. The point interactions are at that defined in the
standard way \cite{AGHH} by means of the generalized boundary
values,
  \begin{eqnarray}
L_0(\psi,\vec a) &:=& \lim_{|\vec x-\vec a|\to 0} {1 \over
\ln|\vec x-\vec a|} \, \psi(\vec x)\,, \nonumber \\ L_1(\psi,\vec
a) &:=& \lim_{|\vec x-\vec a|\to 0} [\psi(\vec x) - L_0(\psi,\vec
a) \ln|\vec x-\vec a|]\,. \nonumber
  \end{eqnarray}
Due to its point character, the Hamiltonian acts as free away of
the interaction support, $(H_N(\alpha,\beta)\psi)(x) =
(-\Delta\psi)(x)$ for $x\not\in Y$, and its domain consists of all
functions $\psi\in W^{2,2}(\R^2 \setminus Y)$ which satisfy the
conditions
$$ L_1(\psi,\vec a)+2\pi\alpha L_0(\psi,\vec a)=0 $$
at any point $\vec a$ from the set $Y$. Since the particle mass
plays no role in the following, we choose the units in such a way
that $2m=1$.


\section{The essential spectrum}

Consider first the essential spectrum of $H_N(\alpha,\beta)$. It
is well known for the so-called straight polymer, i.e.
$H_2(\alpha,\pi)$, which is discussed in \cite[Sec.~III.4]{AGHH}.
In this particular case the spectrum is purely absolutely
continuous and has at most one gap. Specifically, it equals
$[E_0,E_1] \cup [0,\infty)$, where $E_0<E_1<0$, for the coupling
stronger than a critical value, $\alpha<\alpha_Y$, while in the
opposite case the two bands overlap, $E_1\ge 0$, and the spectrum
covers the interval $[E_0,\infty)$. The values $E_0, E_1$, and
$\alpha_Y$ are given as implicit functions of the parameters
$\alpha$ and $l$.
\begin{proposition}
The relation $\inf\sigma_{ess}(H_N(\alpha,\beta)) =
\inf\sigma(H_2(\alpha,\pi))$ holds for any $\beta$ and $N$.
\end{proposition}
{\sl Proof:} The easy part is to check the inclusion
$\sigma_{ess}(H_N(\alpha,\beta)) \supset \sigma(H_2(\alpha,\pi))$.
Given an arbitrary $\lambda\in \sigma(H_2(\alpha,\pi))$ we
construct a sequence $\{\psi_n\}_{n=1}^\infty$ with $\psi_n(x) =
j_n(x)\phi_\lambda(x)$, where $\phi_\lambda$ is a generalized
eigenfunction of $H_2(\alpha,\pi)$ with the energy $\lambda$ and
$j_n\in C_0^\infty(\R^2)$ are mollifier functions to be specified.
If $\mathrm{supp\,} j_n$ intersects just one arm of $Y$, we have
$$ (H_N(\alpha,\beta)-\lambda)\psi_n = (H_2(\alpha,\pi)
-\lambda)\psi_n = -2\nabla\phi_\lambda \cdot \nabla j_n -
\phi_\lambda \Delta j_n\,. $$
Since the functions $\phi_\lambda,\,\nabla\phi_\lambda$ are
bounded, it is sufficient to take $j\in C_0^\infty(\R^2)$ with
$\|j\|=1$ and define
$$ j_n(x) := {1\over n}\, j\left( x-x_n\over n \right) $$
for a suitable sequence $\{x_n\}\subset \R^2$; the latter can be
always chosen in such a way that each $j_n$ intersect with a
single arm of $Y$. Using $\|\nabla j_n\|= n^{-1/2} \|\nabla j\|$
and $\|\Delta j_n\|= n^{-1} \|\Delta j\|$, we conclude that
$(H_N(\alpha,\beta)-\lambda)\psi_n \to 0$ strongly as
$n\to\infty$, i.e. that $\lambda\in \sigma(H_N(\alpha,\beta))$. We
can even choose $\{x_n\}$ so that $\psi_n$ have disjoint supports
forming thus a Weyl sequence, but it is not needed, because
$\sigma(H_2(\alpha,\pi))$ consists of one or two intervals and
$\lambda$ belongs therefore to the essential spectrum of
$H_N(\alpha,\beta)$.

To prove the inequality $\inf\sigma_{ess}(H_N(\alpha,\beta)) \ge
\inf\sigma(H_2(\alpha,\pi))$ we employ the Neumann bracketing. We
decompose the plane into a union
\begin{equation} \label{dec}
 P \cup \left( \bigcup_{j=0}^{N-1} (S_j \cup W_j)
\right), \end{equation}
where $S_j$ is a half-strip centered at the line $\{x\in \R^2: \,
\arg x=\vartheta_j\}$ of the width $d$, $W_j$ is a wedge of angle
$\beta_{j+1}$ between two half-strips $S_j$ and $S_{j+1}$, and
finally, $P$ is the remaining polygon containing the center part
of the ``star''. Introducing the Neumann boundary conditions at
the boundaries, we obtain a lower bound to $H_N(\alpha,\beta)$.
This new operator $\tilde{H}$ is equal to a direct sum of Neumann
Laplacian corresponding to the said decomposition. Since each
wedge part $H_{W_j}$ have a purely continuous spectrum equal to
$\R^+$ and the polygon part $H_P$ has a purely discrete spectrum,
the half-strip parts $H_{S_j}$ are crucial for the threshold of
the essential spectrum of $\tilde{H}$.

We can choose the boundaries such that the distance between the
transverse boundary of the half-strip and the first point
interaction is equal to $l/2$. The spectrum of $H_{S_j}$ on this
half-strip is the same as the symmetric part of spectrum of a
Neumann Laplacian $H_d$ on a ``two-sided'' strip of width $d$,
hence the threshold of $\sigma_{ess}(H_{S_j})$ coincides with that
of $\sigma_{ess}(H_d)$.

Following the standard Floquet-Bloch procedure -- see
\cite[Sec.~III.3]{AGHH} -- we can pass from $H_d$ to an unitarily
equivalent operator which decomposes into a direct integral,
$$ U H_d U^{-1} = {l \over 2\pi} \int_{\theta \in
[-\pi/l,\pi/l)}^\oplus H_d(\theta)\, d\theta, $$
where $H_d(\theta)$ is a point-interaction Hamiltonian in
$L^2([0,l]\times [-d/2,d/2])$ which satisfies the Bloch boundary
conditions,
$$ \psi(l-,y)=e^{i\theta l} \psi(0+,y), \qquad {\partial\psi \over
\partial x}(l-,y)=e^{i\theta l}\, {\partial\psi \over
\partial x}(0+,y) $$
for $y\in [-d/2,d/2]$. The position of the point interaction is
chosen as $(a,0)$. Then it is easy to write the corresponding free
resolvent kernel with one variable fixed at that point,
  \begin{eqnarray}
G^d_0 (\vec x, \vec a; \theta,z) &=& {1 \over l} {2 \over d}
\sum_{m=-\infty}^\infty \sum_{n=0}^\infty {e^{i \left({2\pi m
\over l} +\theta \right) (x-a)} \over \left({2\pi m \over l}
+\theta \right)^2 +\left({\pi n \over d} \right)^2 -z} \nonumber
\\ && \times \cos \left( n \pi {y+d/2 \over d} \right) \cos \left( n\pi
{0+d/2 \over d} \right). \nonumber
  \end{eqnarray}
Using the formula \cite[5.4.5.1]{BMP} one can evaluate the inner
series getting
  \begin{eqnarray}
G^d_0 (\vec x, \vec a; \theta,z) &\!=\!& {1 \over l}
\sum_{m=-\infty}^\infty \, e^{i \left({2\pi m \over l} +\theta
\right) (x-a)} \Biggl\lbrack {1 \over d} \, {1 \over
\varkappa_m^2(\theta,z)} \nonumber \\ && + \, {1 \over
\varkappa_m(\theta,z)} \, {\cosh ((d-|y|)\varkappa_m(\theta,z))
+\cosh(y\varkappa_m(\theta,z)) \over 2
\sinh(d\varkappa_m(\theta,z))} \Biggr\rbrack \nonumber,
  \end{eqnarray}
where $\varkappa_m(\theta,z)=\sqrt{\left( {2\pi m \over l} +\theta
\right)^2 -z}\,$.

To compute the generalized boundary values $L_0$ and $L_1$, and
from them the eigenvalues of $H_d(\theta)$, we follow the
procedure from \cite{EGST}. The coefficient at the singularity
does not depend on the shape of the region \cite{T}, i.e. we have
$L_0(\psi,\vec a)= -1/2\pi\psi(\vec a)$. The value $L_1$ is
expressed by means of the regularized Green's function,
$\xi(\varepsilon;\theta,z):= \lim_{|\vec x-\vec a|\to 0} (G^d_0
(\vec x,\vec a;\theta,z)+ \ln|\vec x-\vec a|/2\pi )$, where we
introduced the $\varepsilon=1/d$ with a later purpose in mind; to
compute it we replace the term $\ln|\vec x-\vec a|$ by its Taylor
series and perform the limit $\vec x\to \vec a$ under the series.

Recall that we are interested in the lowest eigenvalue of
$H_d(0)$, and that due to general principles \cite[Sec,~8.3]{We} a
single point interaction gives rise to at most one eigenvalue in
each gap for a fixed $\theta$, and this is given as a solution to
the implicit equation $\alpha=\xi(\varepsilon,\theta,z)$. Putting
$\theta=0$, we have for $z$ in the lowest gap
  \begin{eqnarray}
\xi(\varepsilon;0,z) &=& \varepsilon \sum_{m=-\infty}^\infty {1
\over \left( {2\pi m \over l} \right)^2-z} \,+\, {1 \over 2 l
\sqrt{-z}} \, {\cosh(\varepsilon^{-1}\sqrt{-z})+1 \over
\sinh(\varepsilon^{-1}\sqrt{-z})} \nonumber \\ && + {1 \over l}
\sum_{m=1}^\infty \left[ {1 \over \varkappa_m(0,z)} \,
{\cosh(\varepsilon^{-1}\varkappa_m(0,z))+1 \over
\sinh(\varepsilon^{-1}\varkappa_m(0,z))} \,-\, {l \over 2\pi m}
\right]. \nonumber
  \end{eqnarray}

So far we made no assumption about $d$. It is obvious that the
decomposition (\ref{dec}) can be chosen in such a way that $d$ is
an arbitrarily large number. For a large $d$, i.e. small
$\varepsilon$, the first two terms of Taylor series for the
eigenvalue read
  $$
z(\varepsilon,\alpha) \,=\, z(0,\alpha) + \left.{\partial z \over
\partial \varepsilon}\right|_{\varepsilon=0} \varepsilon +
\OO(\varepsilon^2)\,,
 $$
where the first derivative is easily computed by means of the
implicit-function theorem,
  $$
\left.{\partial z(\varepsilon,\alpha) \over
\partial \varepsilon}\right|_{\varepsilon=0} = -4 \,
{\sum_{m=-\infty}^\infty \left[ \left( {2\pi m \over l} \right)^2
-z(0,\alpha) \right]^{-1} \over \sum_{m=-\infty}^\infty \left[
\left( {2\pi m \over l} \right)^2 -z(0,\alpha) \right]^{-{3 \over
2}}}\,;
  $$
it is well defined and negative for $z(0,\alpha)<0$.

Hence $z(\varepsilon,\alpha)$ approaches $z(0,\alpha)$ from below
as $\varepsilon\to 0$; it remains to prove that the limit value is
the threshold $E_0$. To this aim, we apply the Floquet-Bloch
decomposition to the operator $H_2(\alpha,\pi)$. The formulae for
Green's function and the $\xi$-function at $\theta=0$ change to
  $$
G_0 (\vec x, \vec a; \theta,z) \,=\, {1 \over 2 \, l}
\sum_{m=-\infty}^\infty {e^{-\varkappa_m(\theta,z)|y-b|} \over
\varkappa_m(\theta,z)} \, e^{i \left({2\pi m \over l} +\theta
\right) (x-a)}
  $$
and
  $$
\xi(0,z) \,=\, {1 \over 2 \, l \sqrt{-z}} \,+\, {1 \over l}
\sum_{m=1}^\infty \left[ {1 \over \varkappa_m(0,z)} \,-\, {l \over
2\pi m} \right],
  $$
respectively. It is obvious that the solution to the equation
$\alpha=\xi(z)$ equals $z(0,\alpha)$. Summing the argument, we
found that $\inf\sigma(H_d)\to \inf\sigma(H_2(\alpha,\pi))$ from
below as $d\to \infty$, and therefore
$\inf\sigma_{ess}(H_N(\alpha,\beta)) \ge
\inf\sigma(H_2(\alpha,\pi))-\eta$ holds for any $\eta>0$, which
concludes the proof. \quad \QED \vspace{1em}

If $\alpha\ge\alpha_Y$ the above proof shows that
$\sigma_{ess}(H_N(\alpha,\beta)) = \sigma(H_2(\alpha,\pi))$, while
in the opposite case one should check also that the two spectra
have the same gap. Since the coincidence of the two spectra is not
important in the following, we are not going to discuss this
question here.


\section{The discrete spectrum}

Our main claim in this paper is that the geometry of the star
graph gives rise to a nontrivial discrete spectrum, and that for
some configurations there are many strongly bound states. We shall
state the result as follows:
\begin{theorem} \label{disc}
(a) $\,\sigma_\mathrm{disc}(H_N(\alpha,\beta)) \ne \emptyset$
unless $N=2$ and $\beta =\pi$. \\ [1mm] (b) Let $N,\,l$, and
$\alpha$ be fixed. For any positive integer $n$ and $c\in\R$ one
can choose the graph geometry (making at least one of the angles
$\beta_j$ small enough) in such a way that the number of
eigenvalues of $H_N(\alpha,\beta)$ (counting multiplicity) below
$c$ is not less than $n$.
\end{theorem}
{\sl Proof:} As we mentioned above, the straight polymer has empty
discrete spectrum. The existence of at least one eigenvalue below
the threshold $E_0$ for $N=2$ and $\beta \neq \pi$ has been
established in \cite{E2}. The part (a) then follows from a simple
auxiliary result. Let $\sigma_\mathrm{disc}(H)= \{E_j:\, E_1\le
E_2\dots\le E_N\, \}$ with $N$ finite or infinite be the discrete
part of the spectrum of a self-adjoint operator $H$. We will say
that $\sigma_\mathrm{disc}(H')\le \sigma_\mathrm{disc}(H)$ for
another self-adjoint operator $H'$ if
$\#\sigma_\mathrm{disc}(H')\ge \#\sigma_\mathrm{disc}(H)$ and
$E'_j \le E_j$ for all $j=1,\dots,N$. We claim that adding an arm
to the graph pushes the discrete spectrum down, adding possibly
other eigenvalues on the top of the shifted eigenvalue set.

\begin{lemma} \label{addleg}
$\;\sigma_\mathrm{disc}(H_N(\alpha,\beta)) \ge
\sigma_\mathrm{disc} (H_{N+1}(\alpha,\tilde\beta))\:$ holds for
any $N$ and angle sequence $\tilde\beta = ( \beta_1, \dots,
\beta_{j-1}, \tilde\beta_j^{(1)}, \tilde\beta_j^{(2)},
\beta_{j+1}, \dots, \beta_{N-1})$ with $\tilde\beta_j^{(1)} +
\tilde\beta_j^{(2)} = \beta_j$.
\end{lemma}
The proof would be easy if the two operators allowed a comparison
in the form sense and the minimax principle could be used. It is
not the case, but one can get the lemma by induction from the
following result. As in \cite{AGHH} we denote by $H_{\alpha,Y}$
the Hamiltonian with point interactions supported by an arbitrary
set $Y=\{y_j\}$; we suppose here that all of them have the same
coupling constant $\alpha$.

\begin{lemma} \label{addpoint}
$\,\sigma_\mathrm{disc}(H_{\alpha,Y'}) \le \sigma_\mathrm{disc}
(H_{\alpha,Y})\,$ for any $Y'=Y\cup\{y'\}$ with $y'\not\in Y$.
\end{lemma}
{\sl Proof:} By\cite[Thm.~I.5.5]{AGHH} $H_{\alpha,Y}$ can be
approximated in the norm-resolvent sense by a family of
Schr\"odinger operators $H^\varepsilon_{\alpha,Y} =-\Delta +
V_\varepsilon$ with squeezed potentials supported in the vicinity
of the points of $Y$, and the same is true for $H_{\alpha,Y'}$.
Each part of the potential can be chosen non-positive, and the
parts corresponding to the points of $Y$ may be the same for both
approximating operators, in which case we have
$H^\varepsilon_{\alpha,Y'} \le H^\varepsilon_{\alpha,Y}$ in the
form sense, or even in the operator one if $V_\varepsilon$ and
$V'_\varepsilon$ are regular enough. By minimax principle we infer
that $\sigma_\mathrm{disc}(H^\varepsilon_{\alpha,Y'}) \le
\sigma_\mathrm{disc} (H^\varepsilon_{\alpha,Y})$ holds for any
$\varepsilon>0$, and the relation persists in the limit
$\varepsilon\to 0$ in view of the norm-resolvent convergence.
\quad \QED \vspace{1em}

\noindent{\sl Proof of Theorem~\ref{disc}, continued:} By
Lemma~\ref{addleg} it is sufficient to prove the part (b) for
$N=2$. We choose then $\beta n < 1/2$ which makes it possible to
draw $n$ circles of radius $R<l/2$ centered at the points $(jl,0),
\; j=1,\ldots,n$. Each of them contains exactly two point
interaction, those placed at $(jl,0)=\vec a_1$ and
$(jl\cos\beta,jl\sin\beta)=\vec a_2$. Their distance is therefore
$a=|\vec a_1\!-\!\vec a_2|= 2 jl\sin (\beta/2)$. By imposing
Dirichlet boundary condition at the circle perimeters, we obtain
an operator estimating $H_N(\alpha,\beta)$ from above.

The proof is now reduced to the spectral problem of the
Hamiltonian $\widetilde{H}(\alpha,a)$ with two point interaction
in a circle with the Dirichlet boundary; it is sufficient to show
that to a given $c$ there is $a_0>0$ such that $\widetilde{H}
(\alpha,a)$ has an eigenvalue $\le c$ for each $a\in(0,a_0)$. This
operator has at most two eigenvalues which are solutions of the
following implicit equation,
  \begin{equation}\label{det}
\det\Lambda(\alpha,\vec a_1,\vec a_2;z) = 0
  \end{equation}
with
$$ \Lambda_{ij}(\alpha,\vec a_1,\vec a_2;z) := \delta_{ij}(\alpha
-\xi(\vec a_i;z)) - (1-\delta_{ij}) \widetilde{G}_0(\vec a_i,\vec
a_j;z)\,, \quad i,j=1,2, $$
where $\widetilde{G}_0$ is the integral kernel of the resolvent
$(\widetilde{H}(\alpha,a) -z)^{-1}$ and $\xi(\vec a_i;z)$ is the
regularized Green's function obtained by removing the logarithmic
singularity  at $a_i$. Due to the rotational symmetry of the
region we have
  \begin{equation}\label{G sym}
\widetilde{G}_0(\vec a_1, \vec x;z) = {1 \over 2\pi} \left(
K_0(\varkappa |\vec x -\vec a_1|) - {K_0(\varkappa R) \over
I_0(\varkappa R)}\, I_0(\varkappa |\vec x -\vec a_1|) \right),
  \end{equation}
where $\varkappa=\sqrt{-z}$. Since it is sufficient to consider
$c\le E_0<0$ we may suppose that $z$ is negative. The first
$\xi$-function is easy to compute,
  \begin{equation} \label{xicirc}
\xi(\vec a_1;z) \,=\, {1 \over 2\pi} \left( \psi(1)-\ln {\varkappa
\over 2} -{K_0(\varkappa R) \over I_0(\varkappa R)} \right).
  \end{equation}
The second one is more difficult, because to express it one would
need to replace the formula (\ref{G sym}) by the Green function
with a general pair of arguments. Instead we employ a simple
Dirichlet bracketing argument similar to that used in \cite{EN}.

Consider three Hamiltonians with a single point interaction placed
at $\vec a_2$: $H_-$, with no restriction in the whole plane,
$H_0$, with the Dirichlet condition at the circle with center
$\vec a_1$ and radius $R$ (the same as for $\widetilde{H}
(\alpha,a)$), and finally $H_+$, with the additional Dirichlet
condition at a circle with center $\vec a_2$ and a radius $R'\le
R-a$. These operators satisfy obviously the inequalities $H_- \leq
H_0 \leq H_+$; since we are interested in comparing the negative
spectra, only the interior parts of the last two operators have to
be considered. The inequalities between the ground-state
eigenvalues of the three operators imply inequalities for
corresponding $\xi$-functions,
$$ \xi_+ (\vec a_2;z) \leq \xi (\vec a_2;z) \leq \xi_- (\vec a_2;z). $$
Both $\xi_\pm$ are known, one from \cite{AGHH}, the other from
(\ref{xicirc}) with changed parameters. In this way we get
  $$
\xi (\vec a_2;z) \,=\, {1 \over 2\pi} \left( \psi(1) -\ln
{\varkappa \over 2} -C(\varkappa) \right)
  $$
with the ``error term'' satisfying $0 \leq C(\varkappa) \leq
{K_0(\varkappa R') \over I_0 (\varkappa R')}$.

One can choose $\varkappa_0>0$ in such a way that $2\pi\alpha
-\psi(1) + \ln {(\varkappa_0/2)}$ is positive. Using the
monotonicity of $u\mapsto {K_0(u) \over I_0(u)}$ and the above
inequalities for $C(\varkappa)$, we get from the condition
(\ref{det}) for $\varkappa>\varkappa_0$ the estimate
  $$
\left( 2\pi\alpha -\psi(1) +\ln {\varkappa \over 2} + {K_0
(\varkappa R') \over I_0(\varkappa R')} \right)^2 \geq
\left( K_0(\varkappa a) - {K_0(\varkappa R) \over I_0 (\varkappa
R)} I_0(\varkappa a) \right)^2.
  $$
We employ further the behavior of the modified Bessel functions
$I_0$ and $K_0$ as $\varkappa a\to 0$, see \cite[9.6.12-13]{BMP}.
For small enough $a$ we can thus choose the positive square root
of the r.h.s. and the condition (\ref{det}) has a solution
satisfying
  $$
\varkappa \geq {2 \over \sqrt{a}} \, \exp\left\{ \psi(1)
-\pi\alpha - {K_0(\varkappa_0 R') \over I_0(\varkappa_0 R')}
\right\} (1+ \OO(a)).
  $$
Since the corresponding eigenvalue is $-\varkappa^2$ and all the
$a$ in the estimating operators can be made simultaneously small
by choosing $\beta$ small enough, the proof is finished. \quad
\QED \vspace{1em}

Notice that for small $a$ the estimating operators have only one
eigenvalue. Considering two point interaction in the whole plane,
we see that the estimate is reasonably good: we have
  $$
\varkappa \leq {2 \over \sqrt{a}} \: e^{\psi(1) -\pi\alpha }\, (1+
\OO(a)).
  $$


\section{Numerical results}

\subsection{The Method}

By \cite[Thm.~III.4.1]{AGHH} the Hamiltonian $H_N(\alpha,\beta)$
can be approximated in the strong resolvent sense by a sequence of
Hamiltonians with point interactions supported by a finite set
$\widetilde{Y} \subset Y$. Hence we get a good approximation of
the spectrum cutting the graph arms to a finite length, large
enough. The most natural choice of subsets is to consider stars
with finite number $M$ of point interaction on each arm (and with
the central point.) This operator $H_N(\alpha,\widetilde{Y})$ has
the essential spectrum equal to $\R_+$ and at most $MN+1$ negative
eigenvalues, due to the presence of point interactions. The lower
eigenvalues, those smaller than $E_0$, converge to the eigenvalues
of $H_N(\alpha,\beta)$ as $M$ increases, while the rest
approximates the negative part of the essential spectrum of
$H_N(\alpha,\beta)$.

The eigenvalues can be obtained as solution to the implicit
equation analogous to (\ref{det}), where
$$\xi(\vec a_i;z)={1 \over 2\pi} \left(\psi(1) -\ln
\left({\varkappa \over 2} \right) \right)$$ $$\quad G_0(\vec
a_i,\vec a_j;z)={1 \over 2\pi}\, K_0(\sqrt{-z}\,|\vec a_i-\vec
a_j|)$$
for $i,j=1,\ldots,(MN+1)$. Once we have an eigenvalue $z_0$ it is
easy to write the appropriate eigenfunction $\varphi(\vec x)$.
From \cite[Sec.~II.1]{AGHH} we know
$$ \varphi(\vec x) = \sum_{j=1}^{MN+1} d_j G_0(\vec x, \vec
a_j;z_0),$$
where $d_j$ are elements of an eigenvector of $\Lambda(\alpha,
\widetilde{Y};z_0)$ corresponding to zero eigenvalue.

\subsection{A broken line, $N=2$}

\begin{figure}[!b]
\begin{center}
\includegraphics[height=6cm, width=8cm]{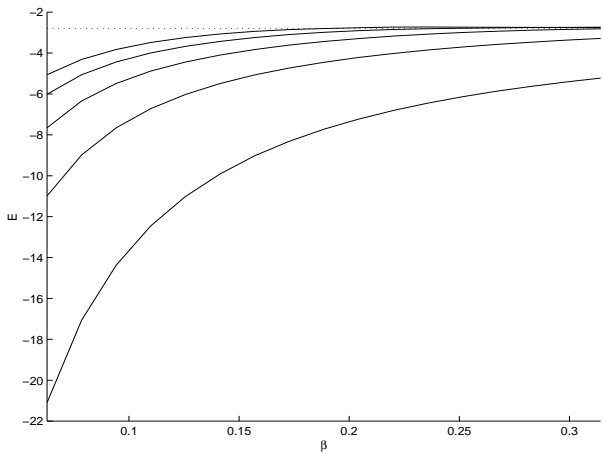}
\end{center}
\vspace{-0.5cm} \caption{Several lower eigenvalues of $H_2(\alpha
=0,\widetilde{Y})$ for small $\beta$. The dotted line is the
threshold $E_0$.} \label{fig:en2 sm}
\end{figure}

\begin{figure}[!t]
\vspace{-0cm}
\begin{center}
\includegraphics[height=6cm, width=8cm]{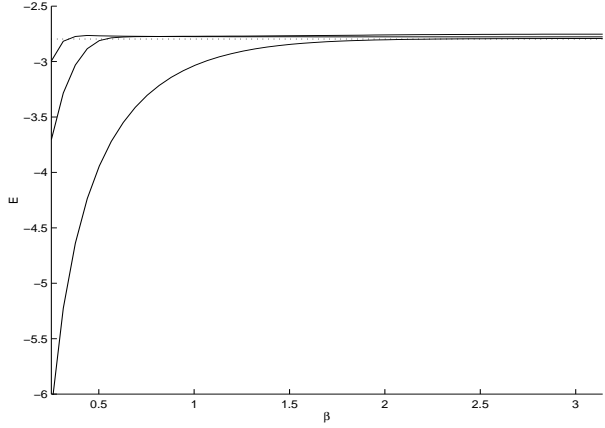}
\end{center}
\vspace{-0.5cm} \caption{The dependence of several lower
eigenvalues of $H_2(\alpha =0,\widetilde{Y})$ on the angle
$\beta$. The dotted line is the threshold $E_0$.} \label{fig:en2}
\end{figure}

Let us start off with a two-arm ``star''. Consider $20$ point
interactions on each arm and $l=1$. We have proven above that
number of eigenvalues of $H_N(\alpha,\beta)$ below a fixed energy
value increases as $\beta$ goes to zero. Numerical results for
$H_2(\alpha=0,\widetilde{Y})$ with $\beta \leq \pi/10$ plotted in
Fig.~\ref{fig:en2 sm} agree with this statement; they also hint
that all eigenvalues are strictly increasing as functions of
$\beta$.
For larger $\beta$, we have a similar situation, see
Fig.~\ref{fig:en2}. We notice that for eigenvalues close to the
threshold $E_0$ the approximation by finite-arm star with $M=20$
becomes insufficient as the picture shows. The eigenvalues above
the threshold will approximate the continuous spectrum as
$M\to\infty$.

\begin{figure}[!b]
\begin{center}
\includegraphics[height=8cm, width=8cm]{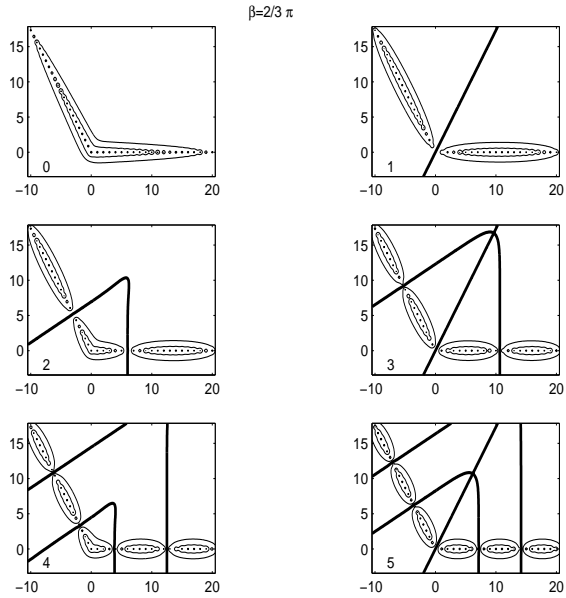}
\end{center}
\vspace{-0.5cm} \caption{Eigenfunctions of six lower states of
$H_2(\alpha=0,\widetilde{Y})$ for $\beta=2/3 \pi$. Only the ground
state has energy below the threshold $E_0$. The bold curves
represent the nodal lines, the contours showing horizontal cuts
correspond to a logarithmic scale.} \label{fig:ef2a}
\end{figure}
\begin{figure}[!t]
\begin{center}
\includegraphics[height=10cm, width=\textwidth]{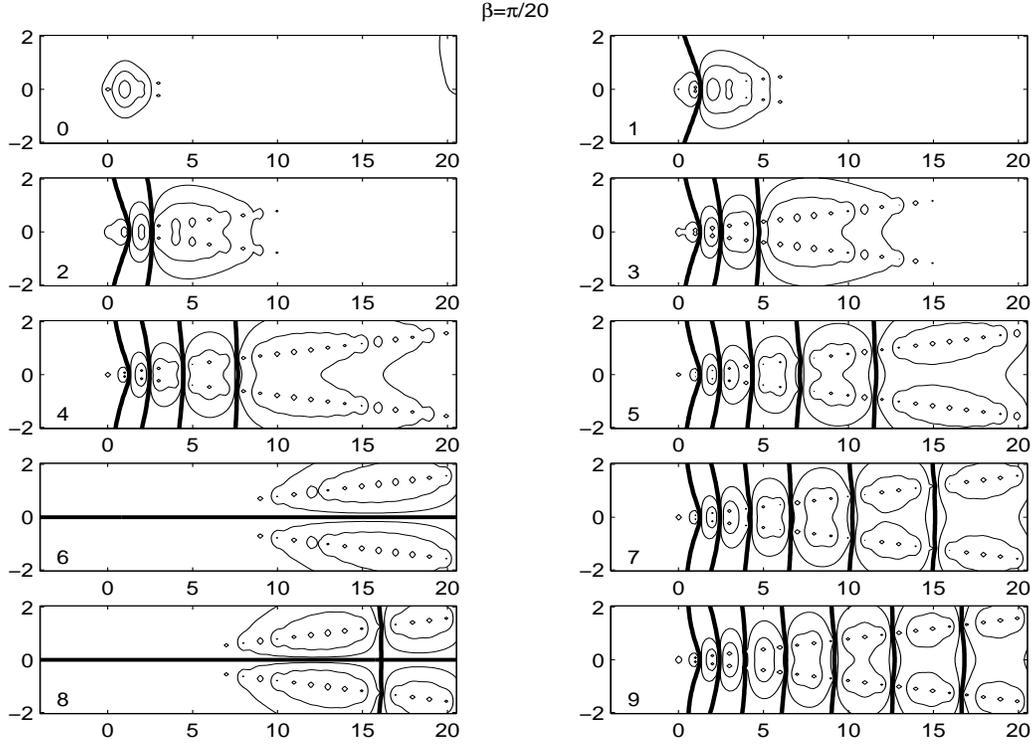}
\end{center}
\vspace{-0.5cm} \caption{Eigenfunctions of ten lowest states of
$H_2(\alpha=0,\widetilde{Y})$ for $\beta=\pi/20$. First five
states correspond to eigenvalues of $H_2(0,\beta)$, the rest would
belong to the essential spectrum in the limit $M\to\infty$. The
bold curves represent the nodal lines, the contours showing
horizontal cuts correspond to a logarithmic scale.}
\label{fig:ef2c}
\end{figure}

In the proof of Theorem~\ref{disc} we observed that pairs of
mutually close point interactions are crucial for the lower part
of discrete spectrum if $\beta$ is small. This behaviour can be
also demonstrated on the corresponding eigenfunctions, compare the
contour graphs in Fig.~\ref{fig:ef2a} to those in
Fig.~\ref{fig:ef2c}. They represent several lowest states of
$H_2(\alpha=0,\widetilde{Y})$ for the angles $\beta=2/3 \pi$ and
$\beta=\pi/20$, respectively. As indicated above, higher
eigenfunction will correspond to the continuous spectrum in the
limit $M \to \infty$. This applies to the eigenfunctions in
Fig.~\ref{fig:ef2a}, except the first one which approximates the
ground state of $H_2(0,2/3\pi)$. In this case it is the only state
which approximates an eigenstate of the infinite star. For a much
smaller $\beta$ in Fig.~\ref{fig:ef2c} there are five eigenvalues
below the threshold (number $0,1,2,3,4$ in the figure), which
correspond to the discrete spectrum of $H_2(0,\pi/20)$. Notice
that the remaining eigenfunctions resemble a standing-wave pattern
along the graph arms as one would expect from an approximation
from a generalized eigenfunction. It may seem that the graph
number $5$ in Fig.~\ref{fig:ef2c} gives rise to a bound state too,
but this only due to an insufficient length $M$ in our
approximation.

\subsection{A three-arm star}

\begin{figure}[!t]
\vspace{-2.5cm}
\begin{center}
\includegraphics[height=13cm, width=12cm]{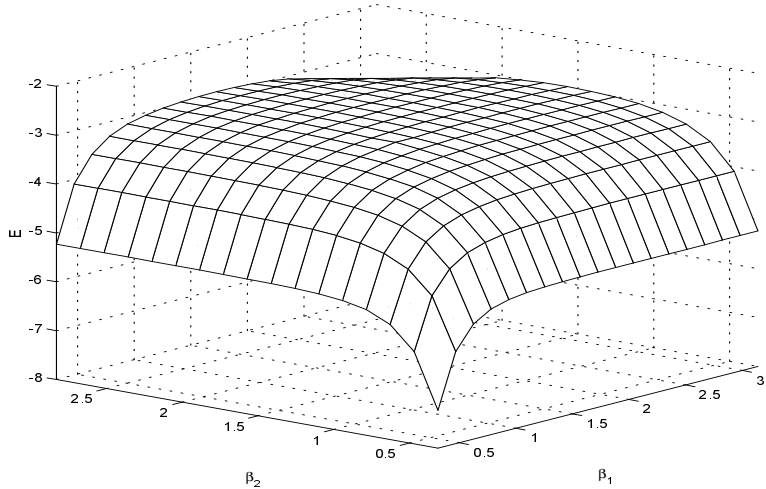}
\end{center}
\vspace{-3.5cm} \caption{Ground state energy of
$H_3(\alpha=0,\widetilde{Y})$.} \label{fig:en3}
\end{figure}

\begin{figure}[!t]
\vspace{0cm}
\begin{center}
\includegraphics[height=7cm, width=12cm]{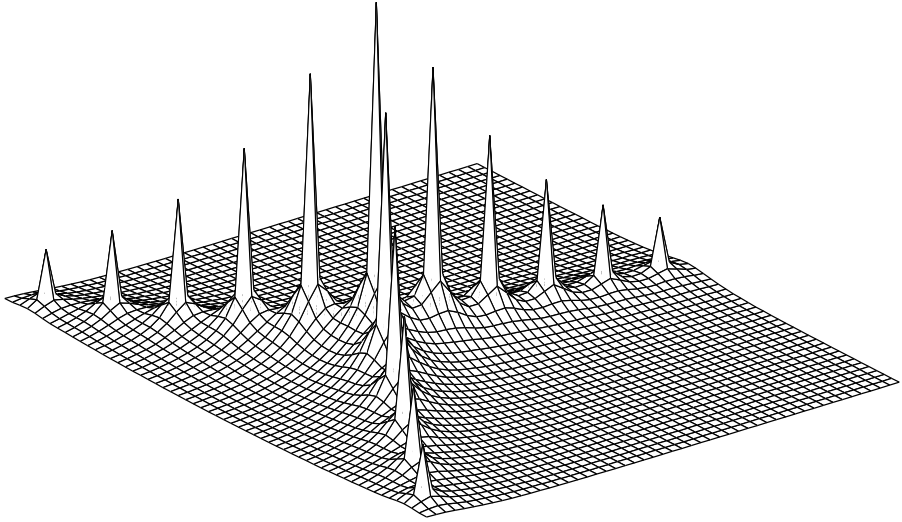}
\end{center}
\vspace{-0.5cm} \caption{Ground state of
$H_3(\alpha=0,\widetilde{Y})$ for $\beta_1=\beta_2=2/3\pi$ which
approximates the ground state of $H_3(0,(2/3\pi,2/3\pi))$.}
\label{fig:ef3}
\end{figure}

Here we consider $10$ point interactions on each arm and we put
$l=1$ again. The behavior of eigenvalues is similar to the two-arm
case, but the spectrum depends of two parameters $\beta_1$ and
$\beta_2$. The minimum binding is achieved in the symmetric case
as the graph of ground state energy of
$H_3(\alpha=0,\widetilde{Y})$ in Fig.~\ref{fig:en3} shows. We see
that the eigenvalue does not change much unless one of the angles
becomes small. The ground state for the symmetric star,
$\beta_1=\beta_2=2/3\pi$, is illustrated in Fig.~\ref{fig:ef3}; we
see the logarithmic singularities at the point-interaction sites
and the overall exponential decay of the eigenfuction along the
graph arms.

\subsection{Larger $N$}
\begin{figure}[!b]
\begin{center}
\includegraphics[height=6cm, width=8cm]{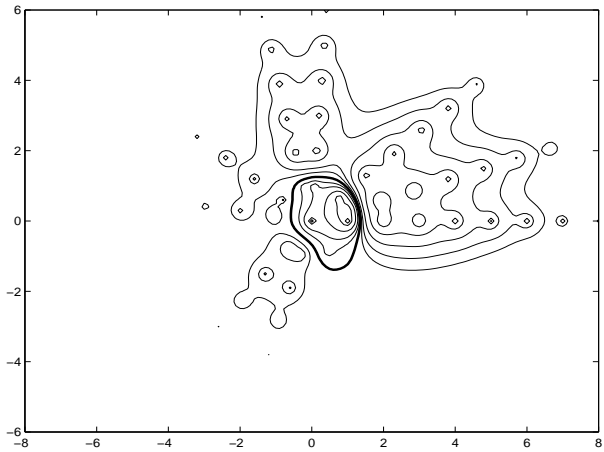}
\end{center}
\vspace{-0cm} \caption{Eigenfunction of third excited state of
$H_{10}(\alpha=0,\widetilde{Y})$ for
$\beta=(0.3,0.7,1.5,1.8,2.5,3,4,4.4,5.2)$. It approximates an
eigenfunction of $H_{10}(0,\beta)$. The bold curve represents the
nodal line, the contours showing horizontal cuts correspond to a
logarithmic scale.} \label{fig:ef10}
\end{figure}
\begin{figure}[!t]
\begin{center}
\includegraphics[height=6cm, width=8cm]{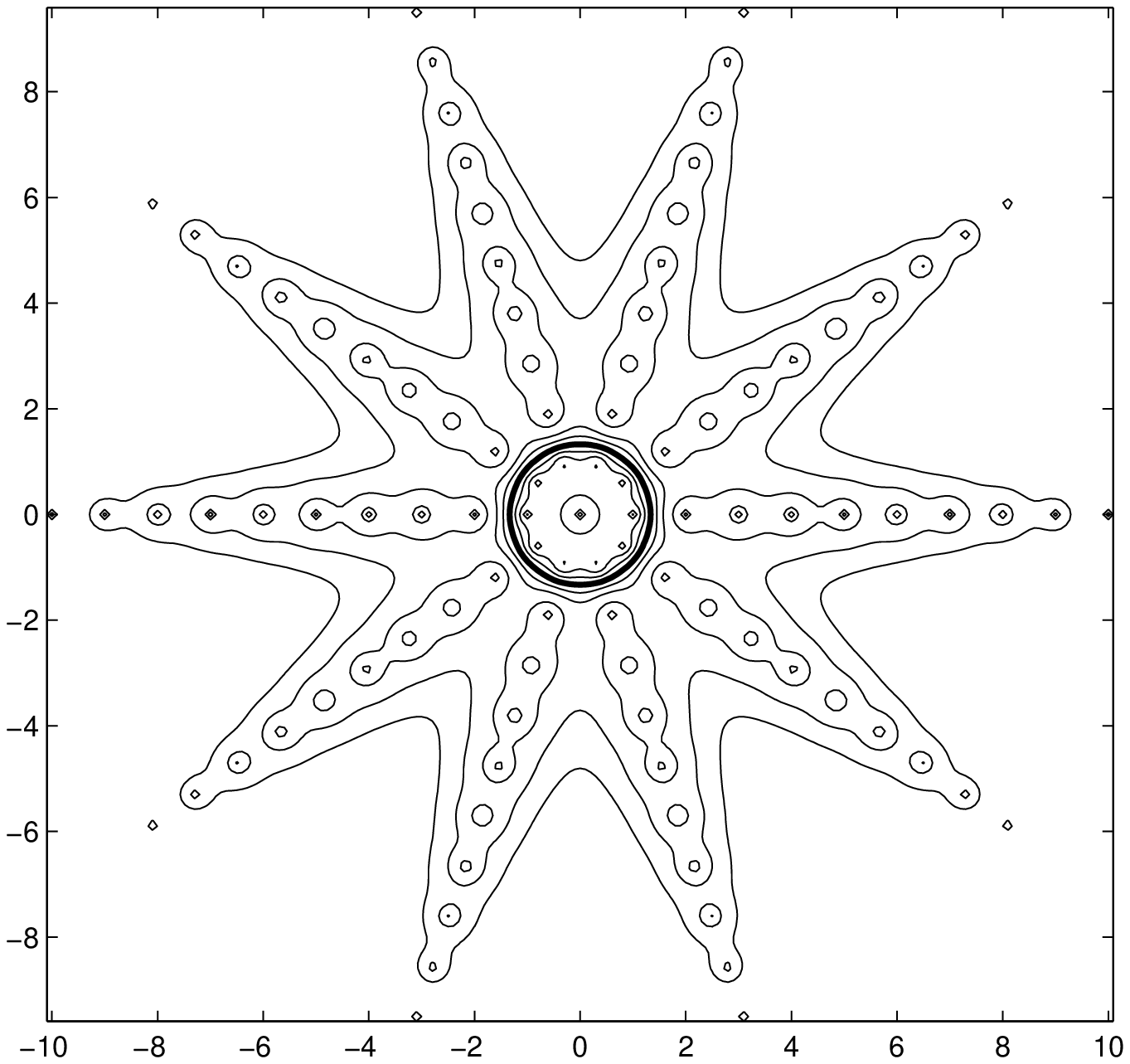}
\end{center}
\vspace{-0.5cm} \caption{Eigenfunction of third excited state of
$H_{10}(\alpha=0,\widetilde{Y})$ for
$\beta=(\pi/5,2/5\pi,\ldots,9/5\pi)$. It approximates an
eigenfunction of $H_{10}(0,\beta)$. The bold curve represents the
nodal line, the contours showing horizontal cuts correspond to a
logarithmic scale.} \label{fig:ef10s}
\end{figure}

In a similar way one can treat star graphs with larger $N$. In
order not to overload the paper with the illustrations, we
restrict ourselves to a single example with $N=10$. The nodal line
plots on the above pictures call to mind the question whether an
eigengunction can have a closed nodal line. Such states can be
found in spectrum of $H_{10}(\alpha,\widetilde{Y})$ for
$\alpha=0$, as it is illustrated in Fig.~\ref{fig:ef10} and
Fig~\ref{fig:ef10s} for a non-symmetric and symmetric star. One of
many mathematical questions which can be asked within the present
model is about the minimum number $N$ for which this is possible.


\subsection*{Acknowledgment}

The authors are grateful to V.~Geyler and K.~Pankrashkin for a
useful discussion. The research was partially supported by GAAS
under the contract \#1048101.

\bibliographystyle{plain}

\end{document}